# Progress in the Next Linear Collider Design*

T.O. Raubenheimer[†]

*Stanford Linear Accelerator Center, Stanford University, Stanford, California 94309 USA*

*Abstract*

An electron/positron linear collider with a center-of-mass energy between 0.5 and 1 TeV would be an important complement to the physics program of the LHC in the next decade. The Next Linear Collider (NLC) is being designed by a US collaboration (FNAL, LBNL, LLNL, and SLAC) which is working closely with the Japanese collaboration that is designing the Japanese Linear Collider (JLC). The NLC main linacs are based on normal conducting 11 GHz rf. This paper will discuss the technical difficulties encountered as well as the many changes that have been made to the NLC design over the last year. These changes include improvements to the X-band rf system as well as modifications to the injector and the beam delivery system. They are based on new conceptual solutions as well as results from the R&D programs which have exceeded initial specifications. The net effect has been to reduce the length of the collider from about 32 km to 25 km and to reduce the number of klystrons and modulators by a factor of two. Together these lead to significant cost savings.

## 1 INTRODUCTION

The Next Linear Collider (NLC) [1, 2] is a future electron/positron collider that is based on copper accelerator structures powered with 11.4 GHz X-band rf. It is designed to begin operation with a center-of-mass energy of 500 GeV or less, depending on the physics interest, and to be adiabatically upgraded to 1 TeV cms with a luminosity in excess of $1 \times 10^{34}\,\text{cm}^{-2}\,\text{s}^{-1}$. The initial construction will include infrastructure to support the full 1 TeV cms to ensure a straightforward upgrade path. A schematic of the NLC is shown in Fig. 1. The collider consists of electron and positron sources, two X-band main linacs, and a beam delivery system to focus the beams to the desired small spot sizes. The facility is roughly 26 km in length and supports two independent interaction regions (IRs).

The NLC proposal was started by SLAC and later joined by LBNL, LLNL, and FNAL. SLAC has formal Memoranda of Understanding (MOUs) with these laboratories and with KEK in Japan to pursue R&D towards a linear collider design. In particular, there has been a close collaboration with KEK for several years concentrated primarily on X-band rf development. The JLC linear collider [3] and the NLC have developed a set of common parameters with very similar rf systems; a status report on the progress of this collaboration was published earlier this year [4]. Work at Fermilab is just starting and will focus on the main linac beam line while the efforts at LBNL and LLNL are focused

*Work supported by the U.S. Department of Energy, Contact Number DE-AC03-76SF00515.

[†] e-mail: tor@slac.stanford.edu

on the damping ring complex, the modulator systems and the gamma-gamma interaction region.

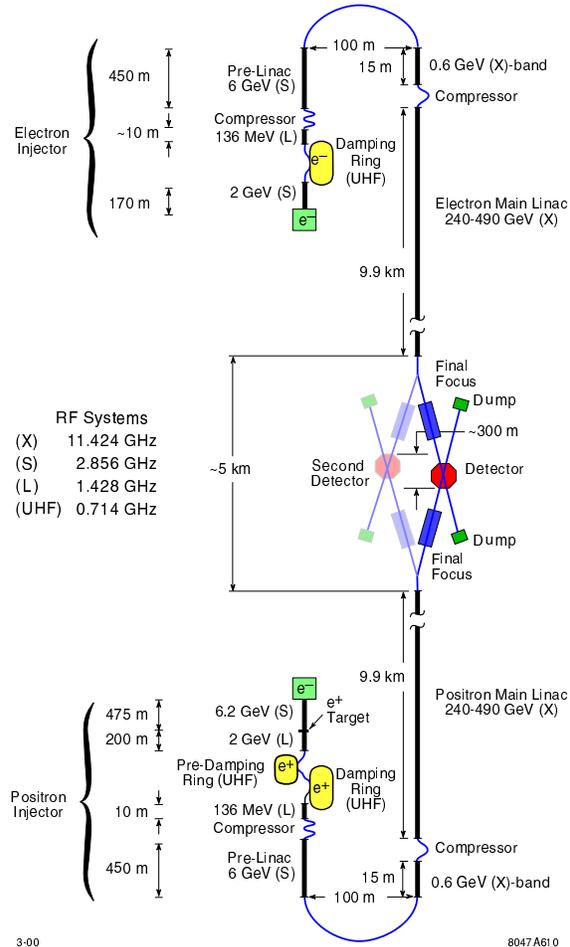

Figure 1: Schematic of the NLC.

In May 1999 for a major DOE review, the NLC project presented both the technical design and a conservative cost estimate for the project. The reviewers concluded that the technical design was in very good shape but questioned the viability of the project with the estimated cost. Over the last year, the NLC collaboration has concentrated on cost reduction and has been able to lower the original estimate by roughly 30%. In addition, the design has been further optimized to meet the physics requirements and there has been continued R&D on key technical components.

In the following, we will first describe recent developments in the NLC rf systems and then discuss the modifications that have been made to the optical design. Next, we will describe some recent modifications to the collider layout that could allow the facility to collide beams with energies as high as 5 TeV once the appropriate rf systems are developed. Finally, we will discuss the NLC luminosity goals and our future plans.

## 2 X-BAND RF SYSTEM

The rf system for the NLC design operates at a frequency of 11.424 GHz to support the higher acceleration gradients needed for TeV-scale colliders. Currently, the NLC rf system is in its third design iteration. The evolution of the rf system has been driven by costing models that have been developed for the collider and by the results from the ongoing R&D programs. The present cost estimate for the rf system has decreased by roughly 50% from that in the 1996 cost model!

The first iteration of the rf system was based on conventional thyratron switched modulators, 50 MW Periodic Permanent Magnet (PPM) focused klystrons, the SLED-II pulse compression system and a Damped-Detuned (DDS) accelerator structure. This configuration was described in the NLC ZDR [1] and is the technology used in the NLC Test Accelerator (NLCTA). The NLCTA began operation in 1997 and verified the beam loading compensation scheme to be used in the NLC as well as the basic rf configuration [5].

The next iteration of the rf design was based on a 75 MW PPM X-band klystron, the Rounded DDS (RDDS) accelerator structure which has 12% higher shunt impedance and the Delay Line Distribution System (DLDS) pulse compression scheme which has significantly higher efficiency than the SLED-II system. This system still used the conventional PFN-type modulators and was presented at the 1999 NLC DOE review; it is described in Ref. [6].

The most recent iteration of the rf design is based on solid-state modulators with an rf pulse length of $3\,\mu$s instead of $1.5\,\mu$s from the klystrons. These parameters reduce the number of klystrons and modulators required by a factor of two. In addition, the rf system uses an enhancement of the DLDS scheme where the rf power is propagated in multiple modes to reduce the amount of waveguide required. In this current design, the rf system for each 250 GeV linac consists of 99 modules each of which contains a modulator, eight 75 MW X-band klystrons, an rf pulse compression unit, and 24 accelerator structures. In the following, we will discuss each of the components in more detail.

### 2.1 Solid State Modulator

The NLC klystrons require roughly 250 Amps at 500 kV. For the 1999 baseline design, the NLC used a conventional PFN-type modulator which would power two klystrons at once. This was a conservative technology choice but it had a maximum efficiency of roughly $65 \sim 70\%$ and it was clearly the most expensive component of the rf system.

Recent improvements in high power Isolated Gate Bipolar Transistor (IGBT) switches have made it possible to consider a solid state modulator design. The switches have relatively fast rise and fall times (<200ns) and can switch a few kA at a few kV [7]. The voltage contributions from a number of switches can be added together inductively in a manner similar to that in an induction linac. The NLC design uses a stack of 80 induction cores, each with two IGBT switches and a 3-turn transformer to generate over 2 kA at 500 kV. This modulator would drive 8 klystrons at once with an estimated cost that is roughly half the cost of the conventional modulator and with an overall efficiency greater than 75%.

In addition to the improved efficiency and reduced cost, the solid-state modulators have a number of other advantages. First, the reliability of the system has the potential to be much higher; failure of a single IGBT should be benign since the core saturates and becomes nearly transparent to the pulse. Additional cores and IGBTs will be included to offset such a loss. Second, the IGBTs will be independently timed to allow for pulse shaping and, for example, offset the natural droop of the pulse as the capacitors discharge.

At this time, a stack of 10 induction cores has been assembled and is being used to power a SLAC 5045 S-band klystron [8]. A full stack of 80 induction cores will be assembled and tested in the fall of 2000.

### 2.2 75 MW PPM X-band Klystrons

Conventional klystrons use a large solenoid magnet to focus the beam between the gun and the collector. Unfortunately, the magnet requires 20 kW of power which is comparable to the average rf output power, effectively decreasing the klystron efficiency. To avoid this a new generation of klystrons using periodic permanent magnet (PPM) focusing have been developed. In these PPM klystrons, the focusing is generated with rings of permanent magnet material which are interleaved with iron pole pieces to generate a periodic axial field.

In 1996, SLAC built a 50 MW PPM klystron which produced $2\,\mu$s long 50 MW pulses with a 55% efficiency. Next, a 75 MW PPM tube was built and was able to produce over 75 MW with a pulse length of $2.8\,\mu$s and an efficiency of roughly 55%, consistent with simulations [9]. At this output level, the pulse length was limited by the modulator output and the repetition rate was limited to 10 Hz due to inadequate cooling of the klystron. A second 75 MW PPM klystron is now being constructed to operate at the full $3\,\mu$s pulse length and 120 Hz repetition rate.

### 2.3 Delay Line Distribution System

The klystrons most efficiently generate a lower power and longer pulse than that needed for the structures. To optimize the system, the rf pulse must be compressed temporally before being sent to the accelerator structures. The SLED-II system, in operation at the NLCTA, compresses the klystron pulse by a factor of 6 but the efficiency is only about 70% so the peak power is only increased by a factor of 4.

To improve on this efficiency, the DLDS system was proposed at KEK [10]. In this system, the power from eight klystrons is summed and divided into equal time intervals. It is then distributed up-beam to eight sets of accelerator structures that are spaced appropriately so that the beam-to-rf arrival time is the same in each case. The power is directed to each different group of structures by varying the relative rf phases of the eight klystrons. The intrinsic efficiency of this system is 100% although wall losses and fabrication errors will likely reduce that to $85 \sim 90\%$.

To reduce the length of waveguide required, a multi-mode version of this system has been developed in which the power is distributed through a single circular waveguide, but in two or more different modes. In the current configuration, each waveguide transports two modes, reducing the length of waveguide by roughly a factor of two. Future studies will investigate both the possibility of transporting four modes in one waveguide and the utility of active rf switching techniques which might allow all the power to be transported in a single waveguide. Finally, to test the components at their design power levels, the NLCTA has been upgraded to produce 240 ns long pulses of 800 MW and testing will begin at the end of FY01.

### 2.4 Accelerator Structures

The accelerator structures for NLC have been studied for many years, much of this in collaboration with KEK. A good summary of the structure development history is given in Ref. [11]. There are three requirements on the structure design: first it must transfer the rf energy to the beam efficiently, second, it must be optimized to reduce the short-range wakefields which depend on the average iris radius, and third, the long-range transverse wakefield must be suppressed to prevent multibunch beam breakup (BBU). The current design for the structures is a 1.8-m traveling wave structure with a filling time of $\sim 100$ ns consisting of 206 separate cells.

To reduce the short-range wakefields, the average iris radius is $a/\lambda \sim 0.18$, leading to a relatively large group velocity ranging from 12% in the front of the structure to 2% at the exit. To optimize the rf efficiency, the structure cells are rounded, improving the shunt impedance by roughly 12% when compared to a simple disk-loaded waveguide like that in the SLAC linac. Finally, the long-range transverse wakefield is suppressed through a combination of detuning the dipole modes and weak damping. The damping is achieved through the addition of four single-moded waveguides (manifolds) that run parallel to the structure and couple to the cells through slots. The signals from this manifold also can be used to determine the beam position with respect to the accelerator structure to micron-level accuracy.

Four of these damped-detuned accelerator structures (DDS) have been built with the most recent structure using rounded cells. Measurements of the rf properties of the structures [12, 13] have confirmed: (1) the cell fabrication techniques which can achieve sub-MHz accuracy, (2) the wakefield models and wakefield suppression techniques, (3) the rf BPMs which are necessary to align the structures to the beam and prevent emittance dilution, and (4) the rf design codes which have sub-MHz accuracy [14].

Although these results are very positive, we have also uncovered a major problem in the structure design. The NLC design calls for a gradient of 70 MV/m to attain a center-of-mass energy of 1 TeV with a reasonable length linac. In the past, we have tested short X-band structures at gradients of over 100 MV/m but it is only recently that we have has sufficient rf power to test the full structures at their design gradient. During these recent tests, damage has been observed after 500 hours of operation. The onset of damage appears to occur at a gradient of $40 \sim 50$ MV/m [15].

The two primary differences between the present structures and those tested earlier at much higher gradients is the structure length and the group velocity of the rf power in the structure. A simple theoretical model has been developed which may explain the correlation with group velocity [16]. Most recently, a workshop was held at SLAC [17] to discuss the breakdown phenomena and the world-wide R&D on high gradient acceleration.

To study the gradient limitation, SLAC and KEK are constructing 12 structures with different group velocities and lengths. In addition, one one of the 1.8-m structures has been cut in two and the last $\frac{1}{3}$ of the structure, where the maximum group velocity is 5%, is being tested. This shortened structure has reached a gradient of 70 MV/m roughly 10 times faster than the full length structures without evidence of rf damage after about 200 hours of operation— a very encouraging initial result! Finally, working on the assumption that the correlation of gradient with group velocity is correct, we are in the process of designing a replacement structure for the NLC so that we can construct this quickly if the testing confirms the initial results. The replacement structure has a lower group velocity which is attained with a phase advance of 150° per cell instead of the standard 120° [18] to keep the average iris radius large.

## 3 OPTICAL DESIGN CHANGES

Over the last year, a number of changes have also been made to the optical design to reduce the collider cost and/or improve the collider performance. In this section, we will discuss two of these changes: the use of permanent magnets and the new design for the beam delivery system (BDS). Other changes include modifications to the bunch compressor system [19], small changes to the beam parameters, possibly placing much of the control electronics directly into the linac tunnels, and modified civil construction techniques to reduce costs.

### 3.1 Permanent Magnets

In the design presented to the 1999 Lehman review, all of the quadrupole magnets, in and downstream of the damping rings, had individual power supplies. This led to an expensive cable plant and a large cost for the redundant power supplies; the redundancy is needed to ensure reliable operation. Given the experience at FNAL with permanent magnets (PM) [20], we have recently been studying replacing the magnets in the injector and the main linacs with variable PM. The desired variation of the quadrupole magnets is 25 ~ 30% which will be sufficient in the main linac and the injector systems, where the optics can be varied, to attain a net operating energy range of 50%. However, because the optics in the BDS is more constrained, we plan to use electromagnets in this region to maintain the full energy flexibility.

The PM have many advantages: they eliminate the cable plant, the redundant power supplies and the cooling systems, the later can also be a source of unwanted vibration. There are roughly 2500 magnets where permanent magnets are being considered as replacements. Presently, there are four different PM designs being studied at FNAL and two prototypes have already been constructed. One of the most significant difficulties in the PM design is the desired stability of the magnetic center as the excitation is changed—the favored beam-based alignment scheme relies on shunting the quadrupoles to determine the offset between the quadrupole magnetic center and the BPM electrical center; changes of the magnetic center with excitation corrupt this measurement. At this time, measurements are being made of the magnetic center stability while alternate beam-based alignment schemes, which would be less sensitive to shifts in the magnetic center, are being studied.

### 3.2 Beam Delivery System

Another significant change to the design is in the beam delivery system (BDS). This region includes the beam collimation section and the final focus. Both of these systems have been completely redesigned over the last year, resulting in a design that is more robust and is half the length of that presented in 1999. The resulting site footprint is roughly 26 km in length rather than 32 km.

The beam collimation system has two purposes: it must collimate the beam tails to prevent backgrounds at the IP and it must protect the downstream components against errant beams. In the previous design, the beam collimation section was designed to survive any mis-steered or off-energy *incoming* beam. This is a difficult constraint because the beam density is normally so high that the beam will damage any material intercepted [21]. The resulting collimation design had to be roughly 2.5 km to collimate 500 GeV beams and the system energy bandwidth was only 1% with very tight optical tolerances—so tight that very small misalignments within the system could cause the beams to damage the beam line components.

In a pulsed linac, the beam energy can change from pulse-to-pulse however large changes to the beam trajectory which are not due to energy errors are much less frequent. We have taken advantage of this fact and redesigned the collimation system to passively survive any off-energy beam but to allow on-energy beams with large betatron errors to damage the collimators. The betatron collimators will be 'consumable' collimators which can be rotated to a new position after being damaged [22]; based on SLC experience, we expect the frequency of the errant betatron errors to be less than 1000 times per year. The net effect of this change in the design specification is that we now have a design that is roughly half as long with much looser tolerances and a larger bandwidth [23].

Another issue that constrains the collimator system design is the wakefields due to the collimators themselves. The collimators are planar devices with very shallow tapers which are expected to minimize the wakefields but make it difficult to perform either direct MAFIA-type or analytic calculations. We have installed a facility to measure these wakefields in the SLAC linac [24]. Initial results show much smaller wakefields than predicted from analytic estimates although the measurements are consistent with MAFIA calculations. We will be using the facility to test additional collimator designs, including some designed at DESY, over the next year.

Second, we have completely redesigned the final focus system (FFS). The previous FFS was based on the lattice of the Final Focus Test Beam (FFTB) at SLAC which was constructed from separate modules for the chromatic correction and made full use of symmetry. Although this makes the design of the FFS simpler, it has the disadvantage of making the FFS quite long—1.8 km for 750 GeV beams.

A new design has been adopted where the chromatic correction of the strong final magnets is performed locally at these magnets [25]. This results in a compact design with many fewer elements which has better performance than the previous version. In particular, the new FFS has a larger energy bandpass with comparable alignment tolerances and a more linear transport which should make it less sensitive to beam tails. Because of the better performance, we have actually increased $L^\star$, the free space from the final magnet to the IP, from 2-m to 4.3-m; this will simplify the design of the interaction region and the interface with the high-energy physics detector.

Finally, the scaling of the length with beam energy in this new design is much weaker than in the earlier design. The present FFS is only 700 m in length but can focus 2.5 TeV beams while an equivalent conventional design would have to be roughly 10 km in length. This change makes it much more reasonable to consider a multi-TeV collider using an advanced high-gradient rf system such as the CLIC design [26]; otherwise the FFS is longer than the linacs. We have taken advantage of this possibility in the NLC design by eliminating the bending between the main linacs and one of the two interaction regions to prevent synchrotron radiation from diluting the emittance of a very high energy beam. Thus, once a high gradient rf system is developed, the NLC could be upgraded to a multi-TeV facility in a cost effective manner, reusing much of the infrastructure and beam line components.

## 4 LUMINOSITY

The NLC has been designed to provide a luminosity greater than $5 \times 10^{33}\,\text{cm}^{-2}\text{s}^{-1}$ at a center-of-mass energy (cms) of 500 GeV and a luminosity in excess of $10 \times 10^{33}\,\text{cm}^{-2}\text{s}^{-1}$ at 1 TeV cms [2]. To ensure this luminosity, the design has a large operating space with numerous margins and overheads. For example, the injector system has been specified to produce roughly 50% more charge than required in the parameter sets. Similarly, the beam emittance dilution budgets, which are in excess of 300%, are based on the component tolerance specifications without consideration of the emittance tuning techniques pioneered at the Stanford Linear Collider (SLC) [27]; at the SLC, the emittance tuning techniques reduced the emittance dilutions by an order-of-magnitude.

At Snowmass '96, we estimated the luminosity that could be expected if all of the collider subsystems performed as specified. This luminosity is roughly a factor of four higher than the 'design' luminosity and is similar to the TESLA luminosity values which are based on similar assumptions. More recently, many of the component prototypes have returned results that are better than the initial specifications. For example, if the rf structure BPMs perform as measured in the DDS3 and RDDS1 structures [12, 13], the emittance dilution due to wakefields would decrease from the allocated 150% to less than 25%. At this time, we are very confident that the collider will exceed the design goals and we will update the parameter sets based on the results of the ongoing R&D programs while maintaining sufficient operational flexibility to ensure that the luminosity goals are met.

## 5 SUMMARY

Over the last year, the NLC collaboration has been focused on new technology developments and design changes to reduce the facility cost. We are making extensive changes to our baseline rf system and to the beam line optics, reducing the collider footprint from 32 km to 26 km while maintaining the energy reach of the facility. We have also uncovered a high gradient limitation in our accelerator structure design and are vigorously investigating solutions—although earlier structure designs have operated at gradients well over 100 MV/m, the present structures are limited to gradients between 40 and 50 MV/m. Finally, we have also modified the collider layout so that it does not preclude upgrading the facility to a multi-TeV collider once an appropriate rf system has been developed.